# Towards a two-dimensional readout of the improved CMS Resistive Plate Chamber with a new front-end electronics

---

**S. Meola,[?,1] A. Samalan[0] ,M. Tytgat[0] ,N. Zaganidis[0] ,G.A. Alves[1] ,F. Marujo[1] ,F. Torres Da Silva De Araujo[2] ,E.M. Da Costa[2] ,D. De Jesus Damiao[2] ,H. Nogima[2] ,A. Santoro[2] ,S. Fonseca De Souza[2] ,A. Aleksandrov[3] ,R. Hadjiiska[3] ,P. Iaydjiev[3] ,M. Rodozov[3] ,M. Shopova[3] ,G. Sultanov[3] ,M. Bonchev[4] ,A. Dimitrov[4] ,L. Litov[4] ,B. Pavlov[4] ,P. Petkov[4] ,A. Petrov[4] ,S.J. Qian [5] ,P. Cao [5 5] ,H. Kou [5 5] ,Z. Liu [5 5] ,J. Song [5 5] ,J. Zhao [5 5] ,C. Bernal[6] ,A. Cabrera[6] ,J. Fraga[6] ,A. Sarkar[6] ,S. Elsayed ,Y. Assran [,] ,M. El Sawy [,] ,M.A. Mahmoud[8] ,Y. Mohammed[8] ,C. Combaret[9] ,M. Gouzevitch[9] ,G. Grenier[9] ,I. Laktineh[9] ,L. Mirabito[9] ,K. Shchablo[9] ,I. Bagaturia[:] ,D. Lomidze[:] ,I. Lomidze[:] ,V. Bhatnagar[;] ,R. Gupta[;] ,P. Kumari[;] ,J. Singh[;] ,V. Amoozegar[<] ,B. Boghrati[<] ,M. Ebraimi[<] ,R. Ghasemi[<] ,M. Mohammadi Najafabadi[<] ,E. Zareian[<] ,M. Abbrescia[=] ,R. Aly[=] ,W. Elmetenawee[=] ,N. De Filippis[=] ,A. Gelmi[=] ,G. Iaselli[=] ,S. Leszki[=] ,F. Loddo[=] ,I. Margjeka[=] ,G. Pugliese[=] ,D. Ramos[=] ,L. Benussi[>] ,S. Bianco[>] ,D. Piccolo[>] ,S. Buontempo[?] ,A. Di Crescenzo[?] ,F. Fienga[?] ,G. De Lellis[?] ,L. Lista[?] ,P. Paolucci[?] ,A. Braghieri[@] ,P. Salvini[@] ,P. Montagna[@@] ,C. Riccardi[@@] ,P. Vitulo[@@] ,B. Francois[A] ,T.J. Kim[A] ,J. Park[A] ,S.Y. Choi[B] ,B. Hong[B] ,K.S. Lee[B] ,J. Goh[C] ,H. Lee[D]**

**,J. Eysermans[E] ,C. Uribe Estrada[E] ,I. Pedraza[E] ,H. Castilla-Valdez[F] ,A. Sanchez-Hernandez[F] ,C.A. Mondragon Herrera[F] ,D.A. Perez Navarro[F] ,G.A. Ayala Sanchez[F] ,S. Carrillo[G] ,E. Vazquez[G] ,A. Radi[H] ,A. Ahmad[I] ,I. Asghar[I] ,H. Hoorani[I] ,S. Muhammad[I] ,M.A. Shah[I] ,I. Crotty[00] on behalf of the CMS Collaboration**

[0]*Ghent University, Dept. of Physics and Astronomy, Proeftuinstraat 86, B-9000 Ghent, Belgium*

[1]*Centro Brasileiro Pesquisas Fisicas, R. Dr. Xavier Sigaud, 150 - Urca, Rio de Janeiro - RJ, 22290-180, Brazil*

[2]*Dep. de Fisica Nuclear e Altas Energias, Instituto de Fisica, Universidade do Estado do Rio de Janeiro, Rua Sao Francisco Xavier, 524, BR - Rio de Janeiro 20559-900, RJ, Brazil*

[3]*Bulgarian Academy of Sciences, Inst. for Nucl. Res. and Nucl. Energy, Tzarigradsko shaussee Boulevard 72, BG-1784 Sofia, Bulgaria.*

[4]*Faculty of Physics, University of Sofia,5 James Bourchier Boulevard, BG-1164 Sofia, Bulgaria.*

[5] *School of Physics, Peking University, Beijing 100871, China.*

[5 5] *Institute of High Energy Physics, UCAS/CAS, Beijing, China.*

[6]*Universidad de Los Andes, Apartado Aereo 4976, Carrera 1E, no. 18A 10, CO-Bogota, Colombia.*

---

1Corresponding author.

*Egyptian Network for High Energy Physics, Academy of Scientific Research and Technology, 101 Kasr El-Einy St. Cairo Egypt.*
*The British University in Egypt (BUE), Elsherouk City, Suez Desert Road, Cairo 11837- P.O. Box 43,Egypt.*
*Suez University, Elsalam City, Suez - Cairo Road, Suez 43522, Egyp*
*Department of Physics, Faculty of Science, Beni-Suef University, Beni-Suef, Egypt*
[8]*Center for High Energy Physics, Faculty of Science, Fayoum University, 63514 El-Fayoum, Egypt.*
[9] *Universite de Lyon, Universite Claude Bernard Lyon 1, CNRS-IN2P3, Institut de Physique Nucleaire de Lyon, Villeurbanne, France.*
[:] *Georgian Technical University, 77 Kostava Str., Tbilisi 0175, Georgia*
[;]*Department of Physics, Panjab University, Chandigarh 160 014, India*
[<]*School of Particles and Accelerators, Institute for Research in Fundamental Sciences (IPM), P.O. Box 19395-5531, Tehran, Iran*
[=]*INFN, Sezione di Bari, Via Orabona 4, IT-70126 Bari, Italy.*
[==]*ENEA, Frascati, Frascati (RM), I-00044, Italy*
[>]*INFN, Laboratori Nazionali di Frascati (LNF), Via Enrico Fermi 40, IT-00044 Frascati, Italy.*
[?]*INFN, Sezione di Napoli, Complesso Univ. Monte S. Angelo, Via Cintia, IT-80126 Napoli, Italy.*
[@]*INFN, Sezione di Pavia, Via Bassi 6, IT-Pavia, Italy.*
[@@]*INFN, Sezione di Pavia and University of Pavia, Via Bassi 6, IT-Pavia, Italy.*
[A] *Hanyang University, 222 Wangsimni-ro, Sageun-dong, Seongdong-gu, Seoul, Republic of Korea.*
[B]*Korea University, Department of Physics, 145 Anam-ro, Seongbuk-gu, Seoul 02841, Republic of Korea.*
[C] *Kyung Hee University, 26 Kyungheedae-ro, Hoegi-dong, Dongdaemun-gu, Seoul, Republic of Korea*
[D]*Sungkyunkwan University, 2066 Seobu-ro, Jangan-gu, Suwon, Gyeonggi-do 16419, Seoul, Republic of Korea*
[E] *Benemerita Universidad Autonoma de Puebla, Puebla, Mexico.*
[F] *Cinvestav, Av. Instituto Politécnico Nacional No. 2508, Colonia San Pedro Zacatenco, CP 07360, Ciudad de Mexico D.F., Mexico.*
[G]*Universidad Iberoamericana, Mexico City, Mexico.*
[H]*Sultan Qaboos University, Al Khoudh,Muscat 123, Oman.*
[I]*National Centre for Physics, Quaid-i-Azam University, Islamabad, Pakistan.*
[00]*Dept. of Physics, Wisconsin University, Madison, WI 53706, United States.*

*E-mail:* sabino.meola@cern.ch

Abstract: As part of the Compact Muon Solenoid experiment Phase-II upgrade program, new Resistive Plate Chambers will be installed in the forward region. High background conditions are expected in this region during the high-luminosity phase of the Large Hadron Collider, therefore an improved RPC design has been proposed with a new front-end electronics to sustain a higher rate capability and better time resolution. A new technology is used in the front-end electronics resulting in very low achievable thresholds of the order of several fC. Crucial in the design of the improved RPC is the capability of a two-dimensional readout in order to improve the spatial resolution, mainly motivated by trigger requirements. In this work, the first performance results towards this two-dimensional readout are presented, based on data taken on a real-size prototype chamber with two embedded orthogonal readout strips. Furthermore, dedicated studies of the muon cluster size as a function of the graphite resistivity are discussed.

## Contents

## 1 Introduction

The first Resistive Plate Chambers (RPC) detectors [1] were developed for cosmic ray experiments, where low rate capability, good time resolution and low cost per unit of area were needed. In order to serve as muon trigger at collider experiments like the Compact Muon Solenoid (CMS) [2], these same features are required, with the addition of high rate capability and good spatial resolution. The RPC system of the CMS experiment [3] at the CERN Large Hadron Collider (LHC) has been designed to efficiently contribute to the muon trigger providing precise measurement of muon momentum and charge, track reconstruction and particle identification up to a pseudorapidity |[| of 1.9. The CMS muon system worked very well at the nominal luminosity of $2 \cdot 10^{34}\,\mathrm{cm^{-2}s^{-1}}$ reached during the LHC Run I and Run II data taking [4] [5]. Due to the large background expected in the High Luminosity LHC (HL-LHC), the RPC detector system needs to be upgraded in order to meet the spatial resolution imposed by trigger requirements and to increase its rate capability, so to be able to work efficiently in high rate environment. The RPC system upgrade will take place during the CMS Phase-II upgrade program [6]. The best way to improve the spatial resolution consists in developing a two-dimensional readout, while a high rate capability implies the developing of an front-end (FE) electronic able to detect signals of the order of few hundreds `V, allowing a drastic reduction of the average charge per count [7]. In fact, the RPC rate capability is mainly limited by the current that can be driven by the high resistivity electrodes and can be improved by modifying the parameters which define the voltage drop on the electrodes. The possible ways to increase the detectable particle flux are:

- Decrease the electrode resistivity;
- Reduce the electrode thickness;
- Reduce the average charge per count.

The average charge per count reduction is the only viable solution to increase the rate capability while operating the detector at fixed current. As a consequence, a very sensitive FE electronics is required. A new improved RPC (iRPC) chamber has been designed by reducing the electrode and gas gap thickness. A full-size prototype of the iRPC double-gap chamber was built for testing purposes under high irradiation. Overall dimensions of the trapezoidal chamber are 58 (100) cm for the small (large) base with a length of 167 cm. The gaps have an electrode width of 1.4 mm, a high pressure laminate thickness of 1.4 mm and a resistivity of $0.9 - 3 \cdot 10^{10}$ ·cm.
A new front-end electronics developed by the INFN Rome Tor Vergata group has been integrated on the PCB strip plane. Each FE can read 8 channels and is equipped with 8 pre-amplifiers (amplification range 0.2 − 0.4 mV/fC) and 2 full-custom ASIC discriminators with 4 channels each. A pull-up system and LVDS transmitters are integrated inside the FEs. The FEs are directly soldered on the PCB to avoid any noise pick-up and the strips are properly terminated on the other end with a resistor. Typical thresholds of the order of 1-20 fC are achievable, resulting in a low charge avalanche operational regime of the RPC chamber, yielding a lower operational voltage, hence suppressing aging effects. The new front-end electronics is radiation-hard up to a total dose of $\sim 1$ Mrad (equivalent to a Non Ionizing Energy Loss of $10^{13}\,\mathrm{n/cm^2}$ ).

## 2 Experimental setup and analysis method

The iRPC prototype was tested at CERN with cosmics. The LVDS FE signals are read by a CAEN TDC module with a resolution of 100 ps. For the longitudinal direction, the total number of strips connected was 16, of which 2 were noisy and one dead. The noisy strips were neglected as located outside the trigger region and uncorrelated with the trigger. The strips width was 0.5 − 1 cm. The longitudinal strips were operating in double gap mode. For the orthogonal direction, the total number of strips connected was 10, of which 2, outside the trigger region, were noisy and have been neglected. In this case the strips were wider and its width was of 5 cm. The orthogonal strips were positioned upon the top gap, therefore they were operating in single gap mode. Data were taken in the overlap region between longitudinal and orthogonal strips, as indicated in Figure 1.

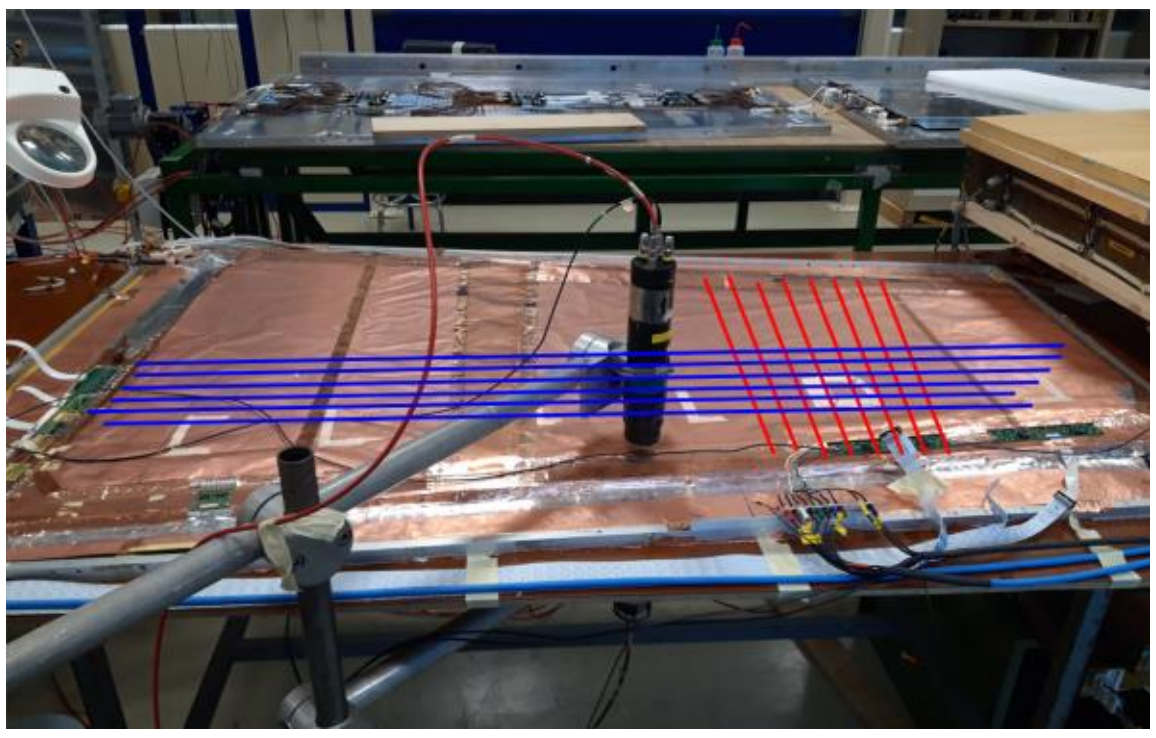

**Figure 1**. The real-size chamber used for the measurements. In blue (red) the longitudinal (orthogonal) strips. Measurements were taken in the strip overlap region.

Hits in the detector recorded by the TDC system are clustered under the following conditions: 1) strips should be adjacent, where the presence of any dead strip is neglected, 2) clustering within a time interval of 10 ns. The time interval of 10 ns is obtained by scanning the time and re-calculating the clusters till a plateau is reached. A systematic uncertainty is taken into account by varying the time interval with 10 ± 4 ns.

The detector efficiency is calculated as the ratio of the amount of hits in the detector over the amount of collected triggers, restricted to the muon time window. The efficiency is measured as a function of high voltage (HV) and in order to extract the necessary parameters, a sigmoid curve is fitted, defined as:

$$\epsilon = \frac{\epsilon_{max}}{1 + e^{-\lambda(HV_{eff} - HV_{50\%})}}$$

where $\lambda$, $HV_{eff}$ and $HV_{50\%}$ are respectively the slope of the curve, the effective high voltage corrected for pressure variations and the high voltage to which corresponds an efficiency of 50% of the maximum efficiency $\epsilon_{max}$. The detector working point is defined as:

$$WP = \ln(19)/\lambda + HV_{50\%} + 150\ \text{V}$$

and is calculated imposing an efficiency of 95% of $\epsilon_{max}$.

## 3 Results of two-dimensional measurements on real-size chamber

An extensive efficiency measurement campaign with 1-dimensional readout have been already performed [8] at the CERN gamma irradiation facility GIF++ [9]. The 1-dimensional measurement of cluster size and muon efficiency for longitudinal and orthogonal strips have been repeated with the new FE electronics for comparison and benchmark using cosmics at the CERN 904 RPC facility.

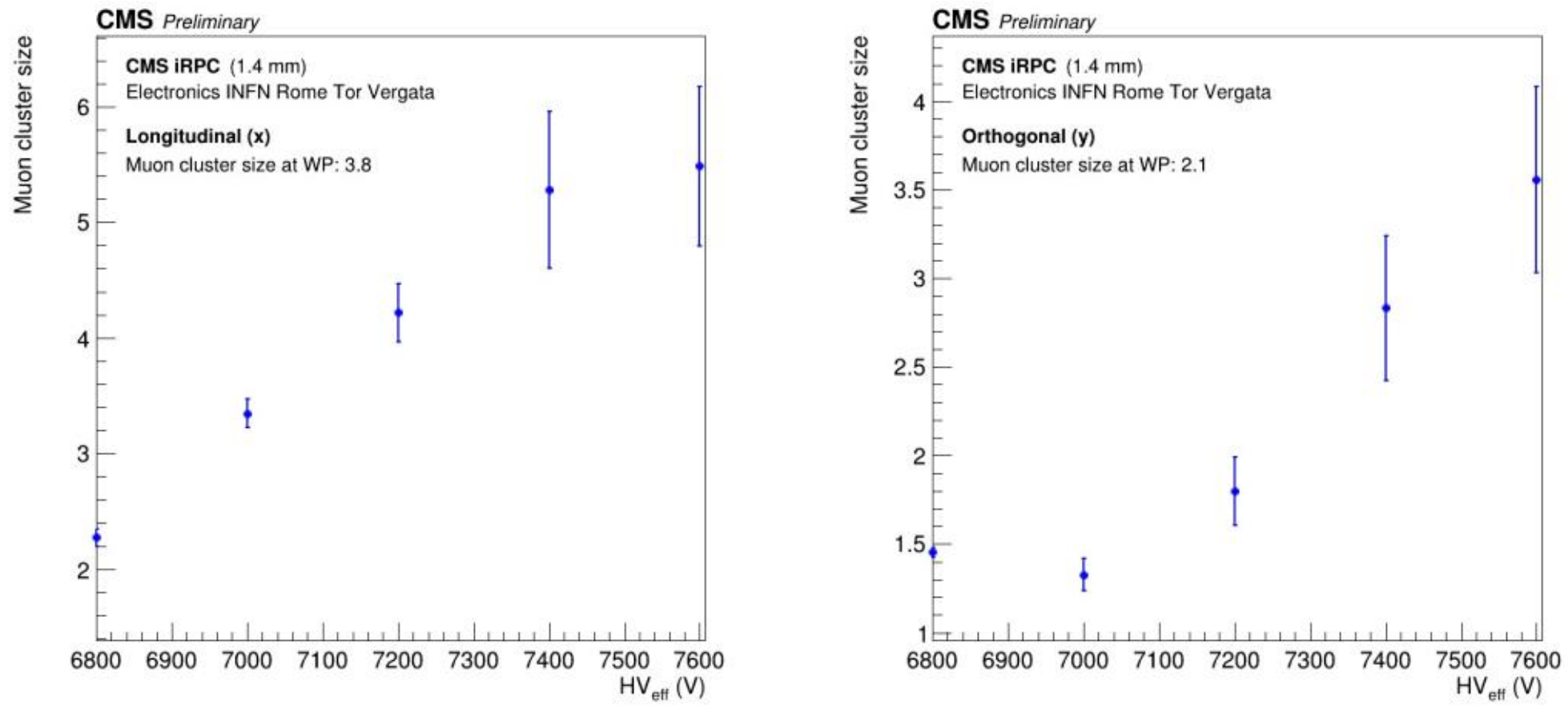

**Figure 2**. Muon cluster size as function of high voltage, on the left for the longitudinal direction and on the right for the othogonal one. The longitudinal (orthogonal) cluster size was measured in the region where the strip pitch is around 0.5 cm (5 cm). The error on the cluster size is estimated by altering the time interval with 10 ± 4 ns.

Figure 2 shows the muon cluster size as a function of the effective high voltage $HV_{eff}$ for the longitudinal and the orthogonal strips, respectively. The smaller cluster size for the orthogonal is due to the larger width of the strips. The error becomes larger at higher voltages as more streamers are present at higher voltages, leading to larger clusters due to separated fired strips in time. Figure 3 shows the muon efficiency for the orthogonal (right) and the longitudinal (left) strips. For the longitudinal strips, a working point of 7.1 kV is measured with an efficiency of 98.7%. The WP is slightly higher than the expected WP for a 1.4 mm iRPC chamber, as the measurement is performed at a long distance from the electronics ($\sim$ 1.5 m), causing a small signal propagation loss along the strip. For the orthogonal strips, a working point of 7.2 kV is measured with an efficiency of 97.1%. A higher working point is expected as the orthogonal strips are on the outer plane of the double gap, therefore sensitive to the induction of charges in one gap. The orthogonal strips also present a lower efficiency due to the fact that the working regime is in single gap mode.

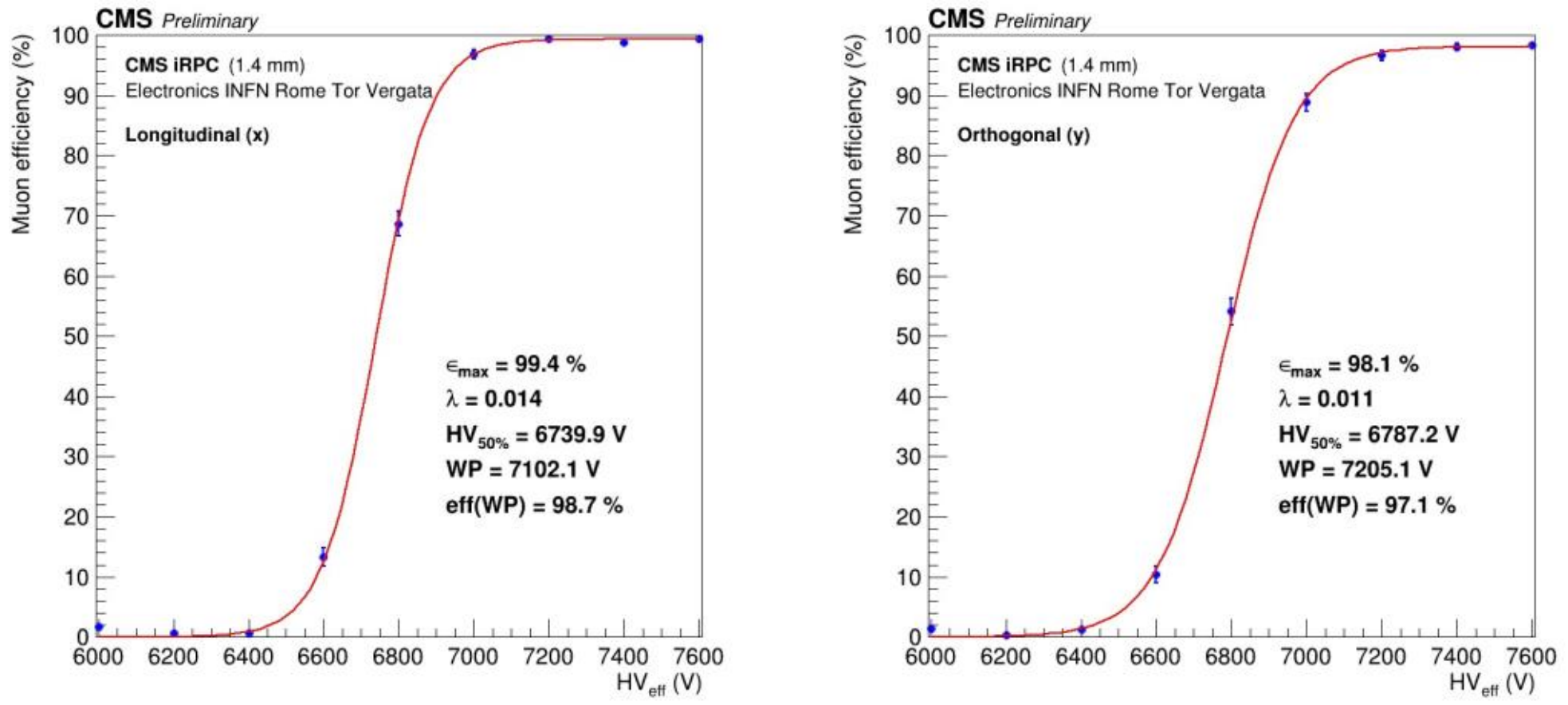

**Figure 3**. Muon efficiency for the orthogonal (right) and the longitudinal (left) strips, as a function of the high voltage.

In Figure 4 is shown the muon efficiency in two-dimensional mode, i.e. for the combination of the longitudinal and orthogonal strips, as function of high voltage. A hit in this configuration requires at least one strip fired simultaneously on both longitudinal and orthogonal direction. The maximum efficiency and working point are driven by the orthogonal strips parameters, working in single gap top mode. The combined efficiency at the working point does not indicate any substantial deterioration in two-dimensional configuration, as shown in Figure 5.

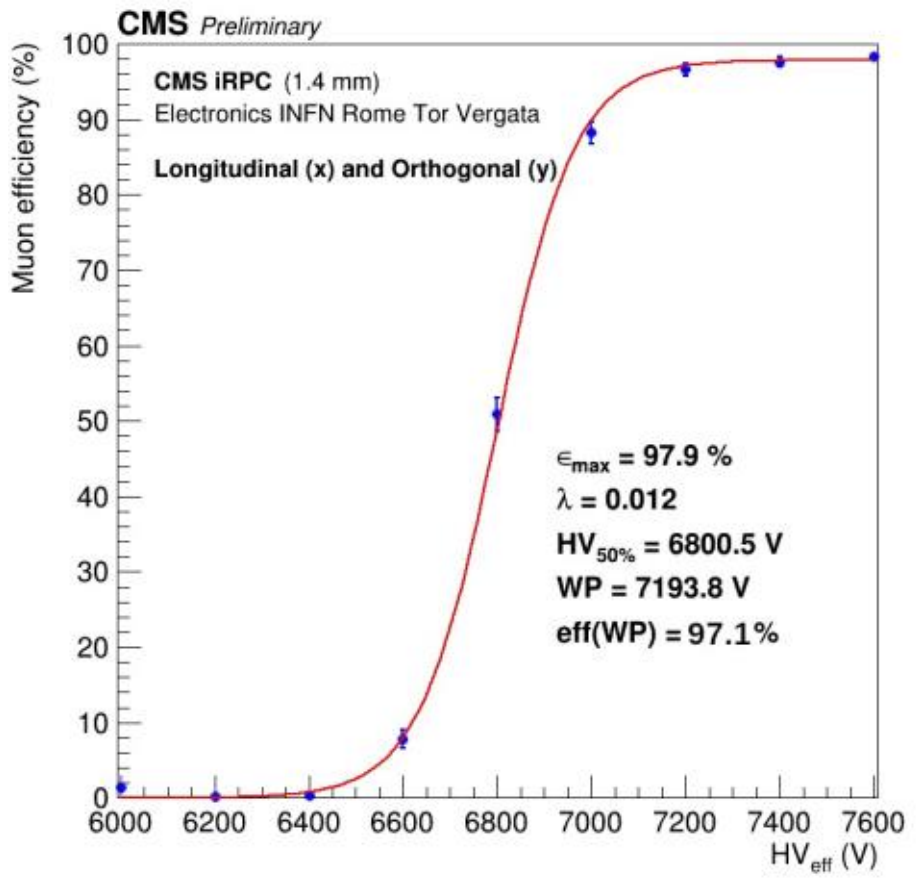

**Figure 4**. Muon efficiency for the combined longitudinal and orthogonal strips as a function of the high voltage.

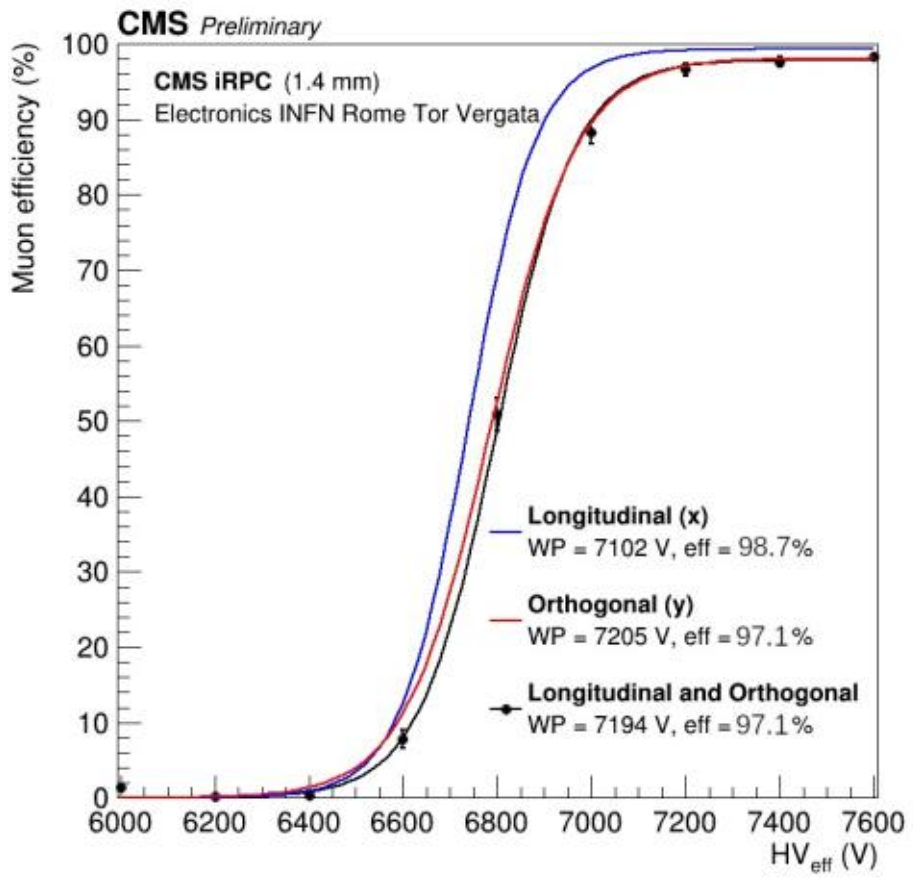

**Figure 5**. Muon efficiency curves comparison between the longitudinal (blue), orthogonal (red) and the combined two-dimensional efficiency (black).

## 4 Conclusion

A real size iRPC 1.4 mm gap equipped with new front-end electronics developed at INFN Tor Vergata has been tested with cosmics. The muon efficiency and cluster size have been evaluated as a function of the high voltage. A combined two-dimensional configuration (Longitudinal + Orthogonal strips) have been considered and efficiency and cluster size measured as a function of the high voltage. The combined maximum efficiency at the working point does not exhibit signs of deterioratiion in the two-dimensional configuration. The results obtained so far show that the considered solution is a suitable alternative for the RPC CMS detector upgrade, more studies at the CERN GIF++ irradiation facility are envisaged to complete the study.

## Acknowledgments

The authors acknowledge the support of the AIDA-2020 project which has received funding from the European Union Horizon 2020 Research and Innovation programme under Grant Agreement No. 654168. Our gratitude also goes to R. Cardarelli, R. Santonico, L. Pizzimento and G. Aielli from Rome Tor Vergata for their support.